\documentstyle[prl,aps,amsfonts,amssymb,epsfig]{revtex}

\begin{document}

\draft

\title{Finite-Size Bosonization of 2-Channel Kondo Model:
a Bridge between Numerical Renormalization Group and Conformal Field Theory}
\author{${}^1$Jan von Delft, ${}^{2,3}$Gergely Zar\'and and
${}^3$Michele Fabrizio}
\address{${}^1$Institut f\"ur Theoretische Festk\"orperphysik, 
Universit\"at   Karlsruhe, 76128 Karlsruhe, Germany \\
${}^{2}$Institute of Physics, Technical University of Budapest,
H 1521 Budafoki \'ut 8., Budapest, Hungary \\
${}^3$International School for Advanced Studies, I-34014, Trieste, Italy}

\twocolumn[\hsize\textwidth\columnwidth\hsize\csname%
@twocolumnfalse\endcsname%
\date{Submitted: June 10, 1997; Revised: February 12, 1998;
Published: Phys.\ Rev.\ Lett., {\bf 81}, 196 (1998).}
\maketitle

\begin{abstract}
We generalize Emery and Kivelson's (EK) bosonization-refermionization
treatment of the 2-channel Kondo model to {\em finite
system size}\/ and on the EK-line analytically construct its exact 
eigenstates and finite-size spectrum. The latter
crosses over to conformal field theory's (CFT) 
universal non-Fermi-liquid spectrum
(and yields  the most-relevant operators' dimensions), 
and further to a Fermi-liquid spectrum in a finite magnetic field.
Our approach  elucidates the relation between bosonization,
 scaling techniques, the numerical renormalization group (NRG) and CFT. 
All CFT's Green's functions are recovered with remarkable ease 
from the model's  scattering states. 
\end{abstract}
\vskip0.0pc]
%\narrowtext
A dynamical quantum impurity interacting with me\-tal\-lic electrons
can cause  strong correlations and sometimes lead to non-Fermi-liquid
(NFL) physics.  A prototypical example is the 2-channel Kondo (2CK)
model, in which a spin-1/2 impurity is ``overscreened'' by
conduction electrons, leaving a non-trivial residual spin object even
in the strong-coupling limit.  Many theoretical treatments of this
model have been developed \cite{CZ97}, including Wilson's numerical
renormalization group (NRG) \cite{CLN80,ALPC} for the crossover from
the free to the NFL regime, Affleck and Ludwig's (AL) conformal field
theory (CFT) \cite{ALPC,AL91} for exact thermodynamic and transport
quantities, valid only near the NFL fixed point, and Emery and
Kivelson's (EK) bosonization-refermionization mapping onto a
resonant-level model \cite{EK92}, valid on a line in parameter space
that connects \cite{Ye} the free and NFL fixed points.  In this Letter
we elucidate the well-known 
\linebreak
yet \mbox{remarkable} fact that these
three approaches, despite tremendous differences in style and
technical detail, yield mutually consistent results: we show that
EK bo\-so\-ni\-za\-tion {\em in a system of finite size $L$}\/ 
yields NRG-like finite-size spectra, and reproduces all
known CFT results.

Our method requires {\em no}\/
knowledge of CFT, only that we bosonize and re\-fer\-mionize with care:
Firstly, we construct the boson fields $\phi$ and Klein factors $F$ in the
bosonization relation $\psi \!  \sim \! F e^{- i \phi}$ explicitly in terms of
the model's original fermion operators $\{ c_{k \alpha j} \}$. Secondly, we
clarify how the Klein factors for EK's refermionized operators act on the
original Fock space.  Thirdly, we keep track of the gluing conditions on all
allowed states.  This enables us (i) to explicitly 
construct the model's finite-size
eigenstates; % $ |\tilde E \rangle$
% explicitly in terms of the $c_{k   \alpha j}$'s;
% illustrating the former's NFL nature; 
(ii) to  analytically obtain NRG-like finite-size spectra that cross over from
free to CFT's universal NFL spectra; 
%(thus directly proving AL's fusion hypothesis) 
%and that  yield the most-relevant operators' dimensions; 
(iii) to describe magnetic-field-induced crossovers exactly;
and (iv) for $L \to  \infty $ to easily  recover all AL's CFT results 
\cite{AL91}. 
%, by showing why all op\-er\-ators can be
%expressed in terms of free boson fields \cite{ML95}. 

{\em The model.---} We consider the standard anisotropic 2CK model
with a linearized energy spectrum\cite {ALPC,AL91,EK92},
\linebreak
defined by $H=H_0 + H_z + H_\perp +  H_h$ ($\hbar=v_F=1$):
\begin{eqnarray}
\nonumber
  H_0 &=& \sum_{k \alpha j} 
  k  : \! c_{k \alpha j}^\dagger  c_{k \alpha j} \! :  , 
\qquad \; H_h = h_i S_z + h_e \hat {\cal N}_s  ,
\\
\nonumber
  H_z + H_\perp &=& \Delta_L
   \sum_{ k k'  \alpha \alpha' j a} 
                  \lambda_a   
  : \! c_{k \alpha j}^\dagger \, 
%  \left( 
\frac12 
\sigma_{\alpha \alpha'}^a S_a 
%  \right)  
   \, c_{k' \alpha' j}  \! \! : \;. 
\end{eqnarray} 
 Here  $c^\dagger_{k \alpha j}$
 creates a free-electron state  $|k \alpha j \rangle$ with
spin 
\linebreak
$\alpha = (\uparrow, \downarrow) $, flavor $j = (1,2)  = 
(+,-)$, radial momentum $ k \equiv   |\vec p| - p_F$,
and normalization 
$
    \{c_{k \alpha j}, c^\dagger_{k' \alpha' j'}\} = 
    \delta_{kk'} \delta_{\alpha \alpha'} \delta_{j j'}
$. 
We let the large-$|k|$ cut-off go to
infinity and  quantize $k$
by defining 1-D  fields  with, for simplicity,
antiperiodic boundary conditions at $x = \pm L/2$ \cite{AL91},
\begin{eqnarray}
  \label{fermionfields}
       && \psi_{\alpha j} (x) \equiv 
        \sqrt \Delta_L %{ 2 \pi \over L}
         \sum_{      n_k \in {\mathbb Z} } 
         e^{-i k x} c_{k {\alpha j}} \; ,  
\end{eqnarray}
where 
 $        k = \Delta_L \left( n_k  - 1/2  \right) $ and
${\Delta_L} \equiv {2 \pi   / L}$ is 
the mean level spacing. By $:\; :$ we denote normal ordering relative
to the Fermi ground state $|\vec 0 \rangle_0$.
 $H_z + H_\perp$ is the Kondo coupling (with dimensionless 
$\lambda_z  \neq  \lambda_\perp   \equiv 
 \lambda_x  \equiv  \lambda_y$) 
to a local spin-1/2 impurity $S_a$ (with
$S_z$-eigenstates $| \!\Uparrow\rangle $, $| \! \Downarrow \rangle$), and
$H_h$ describes magnetic fields $h_i$ and $h_e$ 
coupled to the impurity spin and the total  electron
spin $\hat {\cal N}_s$.

{\em  Conserved quantum numbers.---}
Diagonalizing $H$ requires choosing a suitable  basis.
 Let any  (nonunique) simultaneous eigenstate
of    $
   \hat N_{\alpha j} \equiv
    \sum_{k} \! \! : \! c_{k \alpha j}^\dagger  c_{k \alpha j} \! :
$,
counting the number of $(\alpha j)$ electrons
relative to $|\vec 0 \rangle_0$, be denoted \linebreak
by
$| \vec N \rangle \equiv 
| N_{\uparrow 1} \rangle \otimes 
| N_{\downarrow 1} \rangle \otimes 
| N_{\uparrow 2} \rangle \otimes 
| N_{\downarrow 2} \rangle
$, with 
$ \vec N \! \in \! {\mathbb Z}^4$. 
Since $H$ conserves charge, flavor and total spin,
it is natural to define  new counting operators,
$\hat {\cal N}_y$ ($y = c,s,f,x$), 
\begin{equation}
\label{transformation}
{\textstyle
\left( \begin{array}{c}
        \hat {\cal N}_c \\ \hat  {\cal N}_s \\ 
        \hat {\cal N}_f \\ \hat {\cal N}_x 
        \end{array} \right)
\equiv
{\displaystyle {1\over 2}} \left( \begin{array}{cccc}
        1 & \phantom{-}1 & \phantom{-}1 & \phantom{-}1 \\
        1 & -1  & \phantom{-}1  & -1 \\
        1 &  \phantom{-}1 & -1  & -1 \\
        1 & -1  & -1 & \phantom{-}1 
        \end{array} \right)
\left( \begin{array}{c}
        \hat N_{\uparrow 1} \\ \hat N_{\downarrow 1} 
        \\ \hat N_{\uparrow 2} \\ \hat N_{\downarrow 2}
        \end{array} \right) \; , 
}
\end{equation}
which give the \mbox{total} electron number, the electron
spin, 
flavor, and spin difference between channels, respectively.
Equation~(\ref{transformation}) 
implies that the eigenvalues $ \vec {\cal N}$ are
either all integers or all half-integers (i.e.\ $\vec {\cal N} \in ({\mathbb
  Z} + P/2)^4$, 
with $P = (0,1)$ for  even/odd total electron
number), and that they obey the {\em free gluing condition\vspace*{-1mm}}
\begin{equation}
  \label{gluing1}
  {\cal N}_c \pm  {\cal N}_f = 
 ( {\cal N}_s \pm  {\cal N}_x)\,  \mbox{mod}\, 2 \; \vspace{-1mm}.
\end{equation}
All  non-zero matrix elements of $H_\perp$
have  the form  
%$   \langle {\cal N}_c,  \tilde {\cal N}_s, {\cal N}_f, {\cal N}_x; 
%\Uparrow \! |
%  H_\perp
%  |  {\cal N}_c,  \tilde {\cal N}_s +1 , {\cal N}_f, {\cal N}_x \pm 1; 
%  \Downarrow \rangle$, 
$   \langle {\cal N}_c,  S_T \! - \! 
 {1 \over 2}, {\cal N}_f, {\cal N}_x; \Uparrow \! |
  H_\perp
  |  {\cal N}_c,  S_T \! + \! {1 \over 2} , {\cal N}_f, {\cal N}_x \pm 1; 
  \Downarrow \rangle$, \linebreak
and since the total spin $S_T = \hat {\cal N}_s + S_z$ 
is conserved,
the $\hat {\cal N}_s$-eigenvalue flips only between $S_T \mp {1 \over 2}$,
i.e.\ it fluctuates only ``mildly''. 
In contrast, {\em the $\hat {\cal N}_x$-eigenvalue fluctuates ``wildly''}\/
[an appropriate succession of spin flips 
can produce {\em any} ${\cal N}_x$ 
satisfying (\ref{gluing1})];
this will \linebreak
be seen below to be at the root of the 2CK model's NFL behavior
(in revealing contrast to the 1CK model, which has
no wildly fluctuating quantum number, and  lacks NFL behavior).
For given $( {\cal N}_c, S_T, {\cal N}_f)$
\linebreak
it thus suffices to solve the problem in
the corresponding invariant subspace 
$
   \sum_{\oplus' {\cal N}_x} \!
  | {\cal N}_c, S_T - {1 \over 2} , {\cal N}_f, {\cal N}_x; \Uparrow \rangle 
  \oplus  
  | {\cal N}_c, S_T + {1 \over 2}  , {\cal N}_f, {\cal N}_x
  \! + 1 ; \Downarrow \rangle  ,
$
to be denoted by ${\cal S} $, 
where the prime on the sum indicates its restriction to ${\cal N}_x$-values
respecting (\ref{gluing1}).

{\em Bosonization.---}
To bosonize \cite{EK92} the model in terms
of the original $c_{k \alpha j}$'s \cite{kotliarsi,jvdschoeller},
we define bosonic fields through 
\begin{eqnarray}
\nonumber
\lefteqn{b^\dagger_{q \alpha j} \equiv  
               {i \over \sqrt{n_q} }
               \sum_{n_k \in {\mathbb Z}} 
        c^\dagger_{k+q \alpha j} c_{k \alpha j} , 
        \quad \; ( q \equiv \Delta_L n_q  > 0 ) ,
 }
\\
\nonumber
&&            \phi_{\alpha j} (x) \equiv 
        \sum_{0 <  n_q \in {\mathbb Z}^+}
         \frac{- 1}{ \sqrt{n_q}}
           ( e^{-i q x} b_{q {\alpha j}} 
          +  e^{i q x} b^\dagger_{q {\alpha j}} )
          \, e^{-a q/2},
%          \quad (a \sim {\textstyle {1 \over  p_F}}),
\end{eqnarray}
which account for particle-hole excitations (the
$b$'s by construction satisfy
$
  [ b_{q \alpha j}, b^\dagger_{q' \alpha' j'} ] = 
  \delta_{q q'} \delta_{\alpha \alpha'} \delta_{jj'} 
$ and $[b_{q \alpha j}, \hat N_{\alpha' j'}] = 0$).
Then the usual {\em bosonization relation\vspace*{-1mm}}\/  
\begin{eqnarray}  
 \label{fermionboson}
   &&\psi_{\alpha j} (x) =
a^{-1/2} F_{\alpha j} 
%    ( F_{\alpha j} /  \sqrt a  )
        e^{-i (\hat N_{\alpha j} - 1/2)
        2 \pi x / L} \, \vspace*{-1mm}
        e^{- i \phi_{\alpha j} (x) } \;
\end{eqnarray}
holds as operator identity, where the  {\em Klein factors}
\cite{kotliarsi} $F_{\alpha j} \!\equiv \! \sqrt a \, \psi_{\alpha j} (0) 
e^{i \phi_{\alpha j} (0)}$ 
satisfy (see \cite{jvdschoeller}) 
$[F_{\alpha j}, \hat N_{\alpha' j'} ]\! = \! 
\delta_{\alpha \alpha'} \delta_{j j'} 
F_{\alpha j}$, $[ F, \phi ] \! = \! 0$
 and $\{ F_{\alpha j} , F_{\alpha' j'}^\dagger \} 
\!= \! 2 \delta_{\alpha  \alpha'} \delta_{jj'}$.
Thus $F_{\alpha j}$, $F_{\alpha j}^\dagger$
ladder between the $N_{\alpha j}$,
$N_{\alpha j} \mp 1$ Hilbert spaces without
creating particle-hole excitations, and
ensure proper $\psi, \psi^\dagger$ anticommutation relations.

To exploit the conserved
quantities in the  ${\cal N}_y$ basis, we now 
use  the  transformation (\ref{transformation}) 
to define new bose fields
$b_{q\alpha j} \!  \to \! b_{q y}$ and 
$\phi_{\alpha j} \! \to \!  \varphi_{y}$.
Writing  $H$ in terms of these  [via (\ref{fermionboson})],
only $\varphi_x$ and $\varphi_s$ couple to the impurity \cite{EK92}:
\begin{eqnarray}
         H_0   &=&   \Delta_L \sum_y 
           \frac{1}{2 } 
            \hat {\cal N}_y^2 
        +   \sum_{y, n_q  > 0}
           q \, b^\dagger_{q y} b_{q y} \; , 
% \quad H_K \!= \! H_z \! + \! H_\perp ,}
\label{H0y}
\\
  \label{bosonh}
  H_z 
  &= & \lambda_z   \Delta_L S_z
  \hat {\cal N}_s 
    +  \lambda_z   \Delta_L S_z \sum_{n_q > 0}
  \sqrt{n_q} \,
  i \, (b_{q s} - b^\dagger_{q s} )  \, , 
\\
  H_\perp  &=& 
  { \lambda_\perp \over 2a} \, 
  e^{i \varphi_s (0)} S_- 
   \sum_{j = \pm}  
  F^\dagger_{\uparrow j} F_{\downarrow j} e^{i j \varphi_x (0) }
  + \mbox{H.c.} 
\end{eqnarray}
To eliminate $H_z$,  make the EK \cite{EK92} 
unitary transformation 
$H'  =  U H U^\dagger$, with 
$U (\lambda_z)   \equiv  e^{i \lambda_z S_z \varphi_s (0)}$.
This yields
\newpage
\noindent
  $H_h' =  H_h$, 
$(H_0   +  H_z)'    =   H_0   +   
\lambda_z \Delta_L \hat {\cal N}_s S_z   +   \mbox{const.}$,
$ S'_\pm   =    e^{\pm i \lambda_z \varphi_s (0) } S_\pm$, and 
$\varphi_s$ incurs a phase shift:
\begin{equation}
  \label{newpsis}
U \varphi_s(x) U^\dagger =
\varphi_s (x) - \lambda_z \pi S_z {\rm sgn} (x) \; \equiv  \; 
\tilde \varphi_{s}(x) .
\end{equation}
We henceforth focus on the EK line of fixed $\lambda_z = 1$. 
\linebreak
 Here $\varphi_s$
decouples from $S_\pm$, and 
by (\ref{fermionboson}) and (\ref{newpsis})
the $\psi_{\alpha j}$'s have phase shifts
$\pm \pi/4$. 
Since this  is just the value known for the NFL fixed point
\cite{ALPC,VladZimZaw}, the
$\lambda_\perp$-induced crossover 
between the free and NFL fixed points 
 can be studied 
on  the EK line \cite{Ye} by solving $H'$ by refermionizing.

{\em Refermionization.---} We first have to define Klein factors
for the ${\cal N}_y$ basis. Since an ``off-diagonal'' product
$F^\dagger_{\alpha j} F_{\alpha' j'}$ acting on any state $| \vec N\rangle$
{\em just changes some of its $N_{\alpha j}$ (and hence
${\cal N}_y$) quantum numbers,}\/ we write 
\begin{eqnarray}
\nonumber
    \label{newFs}
   {\cal F}^\dagger_x {\cal F}^\dagger_s \equiv 
   F_{\uparrow 1}^\dagger F_{\downarrow 1} , & \qquad &
 {\cal F}_x   {\cal F}^\dagger_s  
  \equiv  F_{\uparrow 2}^\dagger F_{\downarrow 2} , 
\\
    \label{newFs}
{\cal F}^\dagger_x {\cal F}^\dagger_f
 & \equiv &  F_{\uparrow 1}^\dagger F_{\uparrow 2} \, ,
\end{eqnarray}
thereby defining new Klein factors ${\cal F}_y,{\cal F}_y^\dagger$ satisfying
$[{\cal F}_{y},{\cal N}_{y'}] = \delta_{yy'} {\cal F}_{y}$,
$[ {\cal F}, \varphi ] = 0$ and 
$\{ {\cal F}_y, {\cal F}_{y'}^\dagger \} = 2 \delta_{yy'}$.
Formally, these operators act on an extended Fock space \cite{jvdzarand}
of states with arbitrary 
$\vec {\cal N}   \in    ({\mathbb Z}   +   P/2)^4$.
Its physical subspace contains only those  states 
that obey 
\linebreak
(\ref{gluing1}), and by (\ref{newFs}) it
is closed under the
{\em pairwise}\/ action 
\linebreak
of ${\cal F}_y$'s.
This simple construction for keeping track of ${\cal N}_y$ quantum numbers 
is the main innovation of this Letter. 

Next we define a  {\em pseudofermion} field $\psi_x (x)$ \cite{EK92} by 
\begin{equation}
  \label{eq:psixdefine}
         \psi_x(x)  \equiv
              a^{-1/2}   {\cal F}_{x}  
% ({\cal F}_{x} / \sqrt a)
        e^{-i (\hat {\cal N}_x - 1/2 ) 2 \pi x / L}
        e^{- i \varphi_x  (x) }   ,
\end{equation}
and expand it as $      \sqrt \Delta_L
         \sum_{\bar k} e^{- i {\bar k} x} c_{{\bar k} x} 
$, by
analogy with (\ref{fermionboson}) and (\ref{fermionfields}), which
imply 
$
    \{c_{{\bar k} x}, c^\dagger_{{\bar k}' x}\} = 
    \delta_{{\bar k} {\bar k}'}$.
In the  $c_{k \alpha j}$ basis,
the  $c_{{\bar k} x}$'s create highly non-linear 
combinations of  electron-hole excitations,
as is clear from their 
{\em explicit}\/ definition, via  
 $\varphi_x$ and ${\cal F}_x$,  in terms of the
$c_{k \alpha j}$'s. 
Since  ${\cal N}_x \in  {\mathbb Z} + \frac{P}{2}$, 
we note that 
$\psi_x$ has a  $P$-dependent boundary condition,
implying 
$ 
  {\bar k} = \Delta_L (n_{{\bar k}} - 
  {\textstyle {1- P \over 2}} ) 
$,
and further that %Moreover, from (\ref{eq:psixdefine}) one can 
$
  {\Delta_L } \left(
        {\hat {\cal N}^2_x}/ {2}  +  \sum_{q > 0} 
         n_q b^\dagger_{q x} b_{q x} \right) = H_{0x} + P/8$,
where  $
        H_{0x}  \equiv \sum_{\bar k} {\bar k} 
          :  \! c_{{\bar k} x}^\dagger  c_{{\bar k} x}  \! : \; $
and $: \; :$ means normal ordering of
 $c_{\bar k x}$'s, with  $ \sum_{\bar k}   : \!  c_{{\bar
    k} x}^\dagger c_{{\bar k} x} \!    : \; \equiv   \hat {\cal N}_x  
-    P/2 $.
%$$ 
%     H_{0x}    = {\Delta_L } \biggl[
%        \frac{\hat {\cal N}^2_x}{2} 
%        -  { P \over 8} 
%        +      \sum_{q > 0} 
%          n_q b^\dagger_{q x} b_{q x} \biggr] ,
%$$
We further define the ``local pseudofermion'' $c_d   \equiv  
{\cal F}_s^\dagger S_-$, implying $c_d^\dagger
c_d   = S_z   +   {1\over 2}$.
Eliminating $\hat {\cal N}_s$ in  the subspace ${\cal S}$ using 
$\hat {\cal N}_s = S_T + {1 \over 2}    -  
c_d^\dagger c_d$ we can rewrite $H'$
as $H_{csf}(b_c,b_f,b_s,{\cal N}_c,{\cal N}_f) +  H_x + E_G$, where 
$H_{csf}$ has a  trivial spectrum
and $H_x$ is quadratic:%a simple resonant level Hamiltonian:
\begin{eqnarray}
\nonumber 
   H_x &=&   
  \varepsilon_d c_d^\dagger c_d  + H_{0x}   +   
    \sqrt{   \Delta_L \Gamma } 
    \sum_{\bar k} (c^\dagger_{{\bar k} x}    + c_{{\bar k} x}) 
    (c_d   -   c_d^\dagger) ,
\\
\nonumber
\label{Hquadratic}
E_G &=& \Delta_L [{\textstyle \frac{1}{2 }} 
%\tilde {\cal N} (\tilde {\cal N}_s + 1 )  
( S_T^2 - {\textstyle {1 \over 4}}) +  P/8 ]
- {\textstyle \frac{1}{2 }} h_i  + h_e (S_T + {\textstyle { 1 \over 2}} ) .
\end{eqnarray}
Here $\Gamma \equiv   \lambda^2_\perp /4 a$ and 
$ \varepsilon_d \equiv h_i - h_e$
is the spin flip energy cost. 
As first  noted by EK \cite{EK92}, who derived  $H'$ 
\linebreak 
for $L   \to   \infty$, 
 {\em impurity}\/ properties show 
 NFL behavior since  ``half the pseudofermion'', ($c_d + c_d^\dagger$),
decouples. 

{\em Diagonalizing $H_x$.---} 
To study the NFL behavior  of  {\em electron}\/ properties,
caused by nonconservation of  ${\cal N}_x$, 
we diagonalize $H_x$. First define pseudofermions
having all non-negative energies:
$\alpha_{\bar k} \equiv   \frac{1}{\sqrt2} (c_{{\bar k} x}   + 
c^\dagger_{- {\bar k} x})$ 
and
\linebreak
$\beta_{\bar k} \equiv   \frac{1}{i\sqrt2} (c_{{\bar k} x}   - c^\dagger_{- 
{\bar k} x
  })$ for ${\bar k} > 0$;  if $P   =  1$ then $\alpha_0  \equiv 
c^\dagger_{0 x}$;
and $\alpha_d \equiv c_d$ (or $c_d^\dagger$) for
$\varepsilon_d   >   0$ (or $\le   0$).  
Then the  $\beta_{\bar k}$'s decouple in $H_x$, 
and a Bogoliubov transformation \linebreak
$\tilde \alpha^\dagger_{\varepsilon} = 
\sum_{n = d, \bar k} \sum_{ \nu = \pm}
B_{\varepsilon n \nu} (\alpha_{n}^\dagger + \nu \alpha_{ n})$
yields  %(see \cite{jvdzarand} for the $B$'s)
\cite{jvdzarand}
\begin{eqnarray}
\nonumber  
  & & H_x  = 
    {\varepsilon_d \over 2 } 
    +   \sum_{\varepsilon \ge 0 }
  \varepsilon \left(\tilde \alpha_{\varepsilon}^\dagger 
  \tilde \alpha_{\varepsilon}  - \frac12 \right) 
   +  
  \sum_{{\bar k} >0} {\bar k} 
  \left( \beta_{{\bar k} }^\dagger  \beta_{{\bar k}}
   +  \frac12 \right) ,
\\
\label{eigenenergies}
  & & \qquad
  {4 \pi \Gamma \varepsilon /
  (\varepsilon^2 - \varepsilon_d^2}) \; = \;  
   -    \cot \pi (\varepsilon / \Delta_L - P /2)
        \; .
\end{eqnarray} 
Equation (\ref{eigenenergies}) for the pseudofermion eigenenergies
$\varepsilon$  implies that each  ${\bar k}$ smoothly
evolves into a corresponding 
$\varepsilon(\bar k)$ as $\Gamma$ is
turned on. Since  $\varepsilon ({\bar k})  \simeq 
 {\bar k}   +  
\frac{\Delta_L}{2}$ (or $\simeq  {\bar k}$)  
\linebreak 
for ${\bar k}   \ll$  (\mbox{or} $\gg$)  $\Gamma$,
%we see nicely how the spectrum separates into a 
%strongly- weakly-perturbed low- and 
%high-energy parts of the spectrum emerges naturally as 
%\mbox{$T_K \simeq \Gamma$ \cite{EK92}.}
we  see very nicely that the spectrum's 
low- and high-energy parts 
are strongly and weakly perturbed,  respectively,
with crossover scale 
\mbox{$T_K \simeq \Gamma$ \cite{EK92}.} 

As mentioned above, the pseudofermions act on an extended 
Fock space. To identify which eigenstates $|\tilde E \rangle$
of $H'$ are physical,
note that each has to adiabatically develop, as $\Gamma$ increases from 0,
from some state  obeying the free gluing
condition (\ref{gluing1}). The latter  can be shown 
\cite{jvdzarand} to develop into 
the general gluing condition (GGC) \cite{Hewson}
that 
$\langle \tilde E | \left[ \sum_{\varepsilon \ge 0} 
    \tilde \alpha_{\varepsilon}^\dagger \tilde 
    \alpha_{\varepsilon}
     +       \sum_{{\bar k} >0} \beta_{{\bar k} }^\dagger  \beta_{{\bar k}} 
\right]
    \mbox{mod} 2 | \tilde E \rangle $
%the parity 
%number of $\tilde \alpha_\varepsilon$, \beta_{\bar k}$
%excitations, 
must be equal to  
$   [ {\cal N}_c  +   {\cal N}_f   -  (S_T + {1 \over 2} 
% \tilde  {\cal N}_s   + 1  
    +   {\textstyle \frac{P}{2}}   -   P_d) ] \mbox{mod} \, 2 $,
%$$
%\label{newgluing}
%    {\cal N}_c   \pm  {\cal N}_f   - \left( \tilde  {\cal N}_s   + 1  
%    \pm    \frac{P}{2}   -   P_d \right)
%    =   \left[ \sum_{\varepsilon \ge 0} 
%    \tilde \alpha_{\varepsilon}^\dagger \tilde 
%    \alpha_{\varepsilon}
%     +       \sum_{{\bar k} >0} \beta_{{\bar k} }^\dagger  \beta_{{\bar k}} 
%\right]
%    \mbox{mod} \vspace*{-2mm} 2 ,
%$$
where $P_d   =   0 (1)$ for $ \varepsilon_d   >   0$ ($\le 0$). 
 The GGC and Eqs.~(\ref{eigenenergies}) together constitute an exact
analytical solution of the 2CK model at the EK-line for arbitrary
$\lambda_\perp$, $h_i$ and $h_e$.

{\em Relation to RG methods.---}\/ 
Our exact solution allows us to implement Anderson's ``poor man's scaling''
and Wilson's NRG treatments of the Kondo problem analytically, thus
illustrating the main idea behind both, {\em namely to try to 
uncover the   low-energy physics via an RG transformation.}\/ 
In the first, the RG is generated by reducing (at fixed $L$, usually $=
  \infty$) the bandwidth while adjusting the couplings to keep the dynamical
properties invariant.  Since the cut-off used when bosonizing is $1/a\, (\sim
p_F)$ and  $a$ occurs in $H'$ {\em only}\/ through $\Gamma$, the scaling
equations \cite{Ye} 
\begin{figure}[htbp]
  \begin{center}
        \leavevmode 
    \epsfig{%height=0.7\linewidth,%
width=0.95\linewidth,%
file=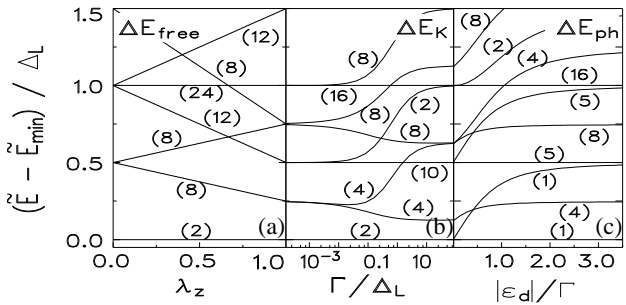}
 \vspace*{1mm}
    \caption{
      All eigenenergies  $\Delta \tilde E   =   
      (\tilde E   -   \tilde E_{min}) / \Delta_L \le 1$ 
       (degeneracies in brackets) 
      of the full $H'$ 
 as functions of 
(a) $\lambda_z \in [0,1]$ at $\Gamma   =   \varepsilon_d   =  0$;
(b) $\Gamma/ \Delta_L \in [0, \infty]$ at $\lambda_z   =   1$,
$\varepsilon_d   =   0$;
(c) $|\varepsilon_d | /\Gamma \in[ 0,3.5]$ at fixed
$\Gamma/ \Delta_L \gg 1$, $\lambda_z   =   1$. 
%Figs.~(b) and (c) were obtained by numerically solving
%      Eq.~(\ref{eigenenergies}),
%see \protect\cite{jvdzarand}\vspace*{-6mm}.
      }\label{fss}
\end{center}
\end{figure}
\noindent
 ${d\ln \lambda_z \over d\ln a}=0$, ${d\ln \lambda_\perp
  \over d\ln a}=1/2$, which imply that $\lambda_\perp$ grows under rescaling
\cite{Coleman}, are exact along the EK-line.  Renormalizing the spin-flip
vertex, possible only approximately in the original $c_{k\alpha j}$ basis by
summing selected diagrams, thus becomes trivial after bosonizing and
refermionizing, which in effect resums {\em all}\/ diagrams into a quadratic
form.

Wilson's NRG \cite{CLN80,ALPC} is, in effect, a finite-size scaling
method which increases (at fixed bandwidth and couplings) the system
size, thus decreasing the mean level spacing and pushing ever more
eigenenergies down into the spectrum's strongly-perturbed regime  below $T_K$.
 Each RG step enlarges the system by 
order $\Lambda   >   1$ by including an extra ``onion-skin shell'' of
electrons, then rescales $H    \to  \Lambda H$ to measure 
energy in units of the
new reduced level spacing. 
%In our case this corresponds to the transformation 
We can mimick this by transforming 
$L   \to    L'   = \Lambda L$ (thus
$\Gamma/\Delta_L   \to   \Lambda \Gamma/ \Delta_L$) and plotting the
spectrum in units of $\Delta_L^\prime   =    {2\pi \over L^\prime}$.

Figure~\ref{fss} displays $(\tilde E   -   \tilde E_{min})/  \Delta_L $
for the lowest few 
\linebreak
$|\tilde E\rangle$ that satisfy the GGC.
Figure~\ref{fss}(a) shows the evolution of
the spectrum {\em toward}\/ the EK-line for
$\lambda_z \in [ 0,1]$ at $\Gamma=$
\linebreak
$\varepsilon_d   =   0$
[i.e.\ 
free fermions, phase-shifted by $\pm \lambda_z \pi/2$ in the spin sector,
see (\ref{newpsis})].  Figure~\ref{fss}(b) shows its further evolution {\em
on}\/ the EK-line for $\Gamma/\Delta_L \in [0, \infty]$ at 
\linebreak
$\lambda_z  
=   1$, $\varepsilon_d=0$.
Decreasing $\Delta_L$ at {\em fixed}\/  $\Gamma$ 
yields an 
\linebreak
NRG-like  crossover spectrum 
that for $\Delta_L   \to    0$ indeed re-
\linebreak
produces the NRG's universal NFL fixed point spectrum  
\linebreak
\cite{CLN80,ALPC} 
(irrespective of the 
specific $\Gamma$ value, illustrating 
the irrelevance of spin anisotropy
\cite{ALPC}).
 This NFL spectrum 
\linebreak
also agrees with that found by AL using 
a so-called {\em fusion
\linebreak
 hypothesis}\/  \cite{AL91},
 which our GGC thus proves simply and directly 
(in contrast to the CFT proof of Ref.~[14(b)]).
Note 
\linebreak
that the ground state (with degeneracy 2) has entropy 
%\linebreak
$\ln 2$, as it must for finite $L$ \cite{entropy}
(in contrast, the celebrated  result $ \frac12 \ln 2$ 
requires taking $L   \to   \infty$ before $T   \to   0$). 

Next we illustrate Wilson's program of extracting the most relevant
operator's dimensions from the $L$ dependence of the finite-size
corrections, $\delta \tilde E(L)   \equiv    {\tilde E}(\Gamma /
\Delta_L)   -   {\tilde E}(\infty) $, to the universal NFL spectrum:
For $\varepsilon_d   =   0$, Eq.~(\ref{eigenenergies}) gives
${\delta\tilde E \over \Delta_L}   \sim  \frac1{\Gamma L}$, thus
{\em on}\/  the EK 
line the least irrelevant operator has dimension 1;
but  perturbative corrections in $\lambda_z   -   1$ yield
${\delta\tilde E \over \Delta_L}   \sim   \frac{\lambda_z   -   1}
{(\Gamma L)^{1/2}}$, thus the general leading irrelevant
operators (absent {\em on}\/ the EK 
line) have dimension $\frac12$
\cite{AL91,Senguptageorge,jvdzarand}.  Next, turning on a local field
$\varepsilon_d   =   h_i $, 
we find from (\ref{eigenenergies}) 
that for $h_i \ll h_c   \equiv    \sqrt{\Gamma/L}  $
%\sim    {1\over L^{1/2}}$ 
the NFL spectrum is only slightly affected,
while 
\linebreak
for $h_c   \ll   h_i   \ll    \Gamma$ 
the spectrum has three distinct regions: 
It is Fermi-liquid-like \cite{ALPC} (with
uniform level spacing) for $\varepsilon \ll h_K   \equiv  
{h^2_i \over \Gamma} $ and  $\varepsilon  \gg    \Gamma$, and
NFL-like (nonuniform level spacings) for $ h_K \ll \varepsilon
\ll \Gamma$. 
 Both the $L$ dependence of $h_c$ and the $h_i$
dependence of the crossover scale $h_K$ show that the local magnetic
field is relevant, with 
\linebreak
dimension $-\frac12$; it causes a crossover,
showsn in Fig.~\ref{fss}(c), to a Fermi-liquid spectrum for all states
with $\varepsilon \ll h_K$.  

For $\Gamma/ \Delta_L   \to   \infty $, $h   \to   0$, we find
logarithmic divergences for the susceptibility, $\chi \approx {1\over
4\pi^2 \Gamma} \ln(\Gamma L)$, and the $\hat {\cal N}_x$
fluctuations, $\langle \hat {\cal N}_x^2 \rangle   \approx   {1
  \over \pi^2} \ln (\Gamma L) $ (with $\langle \hat {\cal N}_x
\rangle= 0$).  
\linebreak
Both are clear signs of 2CK NFL physics: the first
shows that 
no spin singlet is formed due to ``overscreening'', 
the  second how strongly  this
perturbs the electron sea.

{\em Relation to CFT.---}
Recent CFT  \cite{ML95} and scaling \cite{Ye} arguments 
 showed that the NFL regime can be described by 
{\em free boson fields}. 
This can be confirmed very easily
by finding the 
scattering state operators $\tilde c_{{\bar k} x}^\dagger $
[and field $\tilde \psi_x^\dagger (x)$] 
into which the free $c^\dagger_{\bar k x}$'s [$\psi_x^\dagger (x)$]
develop 
when $\Gamma$ is turned on adiabatically as $e^{\eta t} \Gamma$
(at $\varepsilon_d    =   0$),
and deducing from these the behavior of the $\tilde \varphi_y$
fields.
In the continuum limit [$L   \to   \infty$, 
then $(\Delta_L   \ll)$ 
$\eta   \to   0^+$], 
\linebreak
the $\tilde c_{{\bar k} x}^\dagger $'s
 obey \cite{herschfield}  the Lippmann-Schwinger %\vspace{-1mm}
equation
\linebreak
$
  [  H_x , \tilde c^\dagger_{{\bar k} x} ] 
   = {\bar k} \, \tilde c^\dagger_{{\bar k} x}     +   
  i \eta (\tilde c^\dagger_{{\bar k} x}     -   c^\dagger_{{\bar k} x}) ,
$
which gives \cite{herschfield} 
$$
   \tilde c^\dagger_{{\bar k} x} = 
   c^\dagger_{{\bar k} x} + \int { 
      d {\bar k'}\,  2 \Gamma    {\bar k}  \, 
      ( c^\dagger_{{\bar k}' x} +  c_{- {\bar k}' x} )
      \over 
      [    (\bar k    +   i \eta) 
     (\bar k    +   i 4 \pi \Gamma) 
     - \varepsilon_d^2 ] \, 
        ( {\bar k} - {{\bar k}'} + i \eta) }  .
$$
To find the  asymptotic behavior ($|x|   \to   \infty$) of 
$\tilde \psi_x^\dagger (x) \equiv \sqrt {\Delta_L}
\int   d \bar k \, e^{i \bar k x}  \tilde c^\dagger_{{\bar k} x}$,
we may take ${\bar k} / \Gamma \to  0$; this gives 
$$  \tilde \psi_x^\dagger (x) \sim  
1/ \sqrt {\Delta_L}
\int    d \bar k'
  e^{i{\bar k}' x}
  [c^\dagger_{{\bar k}' x} \theta (x) - 
  c_{- {\bar k}' x} \theta (-x)] \; .
$$
Adopting  AL's notation of $L$- and
$R$-movers, $ \tilde \psi_x^\dagger (x) 
  \equiv \theta (x) \tilde \psi_{xL}^\dagger (x)   +  
    \theta (-x) \tilde \psi_{xR}^\dagger (x)$,
this gives $ -  \tilde \psi_{x R}   \sim   \tilde \psi^\dagger_{x L}   \sim
  \psi^\dagger_{x}$. To translate this into ``boundary conditions'' on
the $\tilde \varphi_y$ boson fields, we 
write $\tilde \psi_{xL/R} \equiv \tilde {\cal F}_{x L/R} a^{-1/2} 
e^{- i \tilde \varphi_{xL/R}}$ and note that 
$\tilde \varphi_c$, $\tilde \varphi_f$ 
decouple and $\tilde \varphi_s$ is phase-shifted
as in (\ref{newpsis}). Thus 
the free and scattering boson fields are 
asymptotically related 
(with $ \eta_c, \eta_s, \eta_f   =    1    =   - \eta_x $) by
\begin{equation}
  \label{bosonboundaries}
   (\eta_y \tilde \varphi_{y R}   - \pi S_z \delta_{ys} )\,\sim \,
   (\tilde \varphi_{y L}  
  +  \pi S_z \delta_{ys} ) \, \sim \, \varphi_{y }   \; ,
\end{equation}
while $\eta_y ( \tilde {\cal F}_{yR})^{\eta_y} = 
\tilde {\cal F}_{yL}  =  {\cal F}_{y} $
 for $y = s,f,x$.
This central result, first found in Ref.~\cite{ML95}
(with different phases since Klein factors were neglected), 
shows {\em that the NFL regime can be described by 
boson fields  $\tilde \varphi_{yL/R}$ that are,
asymptotically, free}\/ (with only a trivial $S_z$ dependence).

Next we consider the 16 bilinear fermion currents 
$
    \tilde J^{aA}_y \equiv \,
    : \!   \tilde \psi^\dagger_{\alpha j} 
    (T_y^{aA})_{\alpha \alpha', j j'}  \tilde \psi_{\alpha' j'}  \!   :$
(with
$
T^{00}_c = \frac12 \delta \delta , \;
T^{a0}_s = {\textstyle \frac12} \sigma^a \delta, \;
T^{0A}_f = {\textstyle \frac12} \delta \sigma^A, \;
T^{aA}_x = {\textstyle \frac12} \sigma^a \sigma^A
$),
for which (\ref{bosonboundaries}) 
yields \cite{jvdzarand} 
the ``boundary conditions'' 
$\tilde J^{aA}_{y R}  \sim \eta_y \tilde J^{aA}_{y L}
  $. For $y  =   c,s,f$, these
express the reemergence at the NFL fixed
point of the full $U(1) \times SU(2)_2 \times SU(2)_2$
Kac-Moody symmetry assumed by AL; and for $y  =   x$ 
they are just what AL derived using their fusion hypothesis.
Since these boundary conditions fully 
determine all  AL's 
CFT  Green's   functions \cite{AL91}, the boson approach  will 
{\em identically}   reproduce them too, if one proceeds as follows:
to evaluate $\langle \tilde \psi_{\alpha j} (1) \dots 
\tilde \psi^\dagger_{\alpha' j'} (1') 
\rangle$,  simply insert
(\ref{fermionboson}), rewrite the result in terms of
$\tilde \varphi_{y L/R}$ and $\tilde {\cal F}_{yL/R}$, and  
\mbox{combine (\ref{bosonboundaries}) with 
standard free-boson results like} 
\begin{equation}
  \label{bosoncorrelations}
     { \langle 
      e^{-i \lambda \tilde \varphi_{y R} (x)}
      e^{i \lambda' \tilde \varphi_{y' L} (x')} 
      \rangle 
      \over
      a^{(\lambda^2 + {\lambda'}^2)/2 }  }
          \; \sim \; { \delta_{yy'}
        L^{- {1/2} ( \eta_y \lambda - \lambda')^2}  
        \over (i x - i x')^{\eta_y \lambda \lambda'} }  .
\end{equation}
{\em All  asymptotic NFL behavior of electron Green's func-}
\newpage
\noindent
{\em tions arises from 
the fact that $\eta_x   =   -1$}, combined with relations like 
(\ref{bosoncorrelations}); it directly yields, e.g., 
the so-called ``unitarity paradox'' \cite{ML95}
$\langle \tilde \psi_{\alpha j R}  (x)
\tilde \psi^\dagger_{\alpha' j' L} (x')  \rangle \sim 0$
(for  $L  \to$ 
$   \infty$, then $|x'   -   x |   \to    \infty$). 
Note,  though, that 
probability is 
\linebreak
not lost during scattering: 
$\tilde \psi_x^\dagger (x)$ shows that each
 pseudo\-{\em particle}\/
$c^\dagger_{{\bar k}' x}$ incident from
$x   >   0$ is ``Andreev-scattered'', emerging  at $x   <  0$ 
as pseudo{\em hole}\/  $c_{- {\bar k}' x}$,
{\em orthogonal to what was incident};
 this very NFL-like behavior  dramatically illustrates
the effects of $\hat {\cal N}_x$  nonconservation.

To find AL's {\em boundary operators}\/ in terms of the
$\tilde \varphi_y$'s  \cite{Ye,jvdzarand}, one calculates the
operator product expansion of 
$\tilde \psi_{R \alpha j}
\tilde \psi^\dagger_{L \alpha' j'}$. 
Since $\eta_x   =   -1$, {\em all}\/ terms contain  
a factor $e^{ \pm i \varphi_y}$ ($y   =  s,f$ or $x$)
 with dimension  $\frac12$; this  ultimately 
causes the famous $T^{1/2}$ in the resistivity 
\cite{AL91,Ye,ML95}.

In conclusion, finite-size bosonization
allows one  (i) to mimick, in an {\em exact} way,
the strategy of standard RG approaches, and (ii)
to recover with remarkable ease all exact 
results known from CFT  for the NFL fixed point. It 
thus constitutes a bridge  between  these theories.

We thank T. Costi, C. Dobler, G. Kotliar, A. Rosch, A. Ruckenstein,
A. Sengupta, 
 and particularly  
A. Ludwig, who anticipated some  of our results, for discussions,
A. Schiller for 
showing us how to get $\tilde c^\dagger_{\bar k x}$
(unpublished, 1995), and H. Schoeller for teaching us to 
 bosonize carefully.  G.Z.\ was supported by the Magyary
Zolt\'an Scholarship and the Hungarian Grants OTKA T-021228 
and OTKA F016604, J.v.D.\ by ``SFB~195'' of the 
DFG. 

\vspace{-.8cm}.

\end{document}